# Direct Measurement of the Structural Change Associated with Amorphous Solidification using Static Scattering of Coherent Radiation

Charlotte F. Petersen[1,2] and Peter Harrowell[1,*]

[1]*School of Chemistry, University of Sydney, Sydney, New South Wales 2006 Australia*
[2]*School of Chemistry, University of Melbourne, Melbourne, Victoria 3010 Australia*
[*] Corresponding author: peter.harrowell@sydney.edu.au

Abstract

In this paper we demonstrate that the weak temperature dependence of the structure factor of supercooled liquids, a defining feature of the glass transition, is a consequence of the averaging of the scattering intensity due to angular averaging. We show that the speckle at individual wavevectors, calculated from a simulated glass former, exhibits a Debye-Waller factor with sufficiently large temperature dependence in the to represent a structural order parameter capable of distinguishing liquid from glass. We also extract from the speckle intensities a quantity proportional to the variance of the local restraint, i.e. a direct experimental measure of the amplitude of structural heterogeneity.

# 1. Introduction

In a glass transition, a fluid transforms into an amorphous solid when supercooled with little significant change in the scattering structure factor [1]. The conventional response to the challenge of describing the passage from fluid to solid in the absence of structural change is to eschew a microscopic description entirely and, instead, use the shear viscosity to measure solidification. This dynamic definition of solidification is awkward in that it is unsuitable for the characterization of the low temperature solid whose properties, including rigidity, arise from structural, rather than dynamic, origins. The subsequent search for useful structural signatures of the transformation of liquid to glass in simulations studies has generated a considerable literature [2-10].

Just what it is we mean by 'structural' is important in the following analysis so we shall take a moment to clarify our terminology. In simulations, we define any quantity obtainable from an instantaneous configuration to be *structural* in character, while any property that requires information from trajectories of configurations – generated by integrated equations of motion or stochastically – is *dynamical*.

This generalization of the definition of structural information is important, we argue, because traditional geometric or topological descriptions are of limited use given the multiplicity of structures in a glass [11]. Recently [12-14], we introduced a computational measure of the capacity of a configuration to restrain the motion of individual particles – a measured based on the normal mode analysis of instantaneous configurations – and demonstrated that this structural description provided interesting insights into liquid fragility, the influence of finite size and concentration of pinning sites [12-14] and the loss of restraint at the glass surface associated with observed enhanced kinetics [15]. As the term *glass transition* is defined in terms of dynamical properties, we shall refer to a structural description of the transformation of the supercooled liquid as *amorphous solidification.* The term 'amorphous solidification' does not imply any underlying thermodynamic singularity. All we assume in the following analysis is that the low temperature state is mechanically stable.

Motivated by the success of the restraint order parameter from simulations, we shall explore here the possibility of measuring an analogous structural quantity experimentally. First, however, we need to revisit our definition of 'structural'. Measurements take place over a non-zero time interval. This means there is no direct analog of the instantaneous property accessible to simulations. Experimentally, the term 'structural' is generally associated with the measurement of static scattering intensities. X-ray crystallography, the archetype of structural determination, is based on this type of measurement. Static scattering is characterized by the absence of resolution of the energy of the scattered intensity. On this basis, we shall refer to a property obtainable from a static scattering experiment to be structural in character even through it involves contributions from relaxation dynamics sampled over the measurement time. Quantities that require an energy or time resolved experiment are regarded as dynamical. We shall return to this operational definition of 'structural' in the Section 5.

The order parameter we introduced in refs. [12-14] is an instantaneous analog of the Debye-Waller (DW) factor. For crystals, the DW factor is routinely obtained from static scattering measurements and so, as discussed above, can be regarded as structural information – not a measure of the mean particle positions themselves but a measure of the variance of these positions. In 1984, Tarazona [16] introduced the variance of atomic displacement as an order parameter to describe the freezing of a liquid. The DW factor corresponds to the experimental measure of Tarazona's structural order parameter.

Our goal is to develop an expression for the DW factor for amorphous materials that parallels the analogous crystalline quantity. As discussed in Section 2, the literature on the amorphous DW factor is based on a definition that involves data from inelastic scattering measurements, leaving open the question as to whether this quantity can be obtained from a static scattering experiment. The goal of this paper is to resolve this question by establishing, using simulated scattering data, that the Debye-Waller factor for an amorphous material can be obtained directly from static scattering measurements. We emphasize that the DW factor for an amorphous material is not a new quantity; what is new is the demonstration that the amorphous DW factor can be obtained in a manner analogous to that used to measure the crystal DW factor.

This paper is organized as follows. In the following Section we provide a brief background on the Debye-Waller factor in crystal and amorphous solids. In Section 3 we consider the temperature dependence of static scattering intensities and in Section 4 we present results from the analysis of the temperature dependence of speckle pattern from a coherent radiation source.

**2. Debye-Waller factors in Crystals and Amorphous Solids**

The Debye-Waller factor was originally formulated to account for the reduction of the elastic scattering intensity at Bragg peaks of a crystal due to the disorder introduced by thermally excited vibrations [17]. This reduction is conventionally expressed as

$$\frac{I(\vec{q},T)}{I(\vec{q},0)} = \exp(-2W(\vec{q},T)) \quad (1)$$

where the Debye-Waller (DW) factor, $D_w$, is

$$D_w(\vec{q},T) = \exp(-2W(\vec{q},T)) \quad (2)$$

If we assume that the fluctuations in atomic positions are small enough to be characterized as harmonic, then we can write [18]

$$W(\vec{q},T) \approx \frac{1}{2}\left\langle (\vec{q}\cdot\vec{u}_i)^2 \right\rangle_{eq} \quad (3)$$

where $\vec{u}_i$ is the displacement of atom *i* from its equilibrium position and the average $\langle\ldots\rangle_{eq}$ is the equilibrium average over the harmonic Hamiltonian and the atoms in the system. In the context of X-ray crystallography, the DW factors are, unequivocally, structural measures, complimentary to the atomic positions. They or the associated mean squared displacement of atoms are standard outputs of structural determinations by X-ray diffraction [19]. Nonlinear least square fitting of the structure factors, such as the Rietveld method [20], treat both the mean atomic positions and the mean squared amplitudes atomic displacements as adjustable parameters.

Turning to amorphous materials, the Debye-Waller factor has been obtained from experiments that explicitly include inelastic processes. Buchenau and coworkers [21,22] have defined the DW factor for an amorphous material in terms of the inelastic neutron scattering spectrum $S(\vec{q},\omega,T)$ as follows

$$D_w(q,T) = \frac{S_{el}(q,\Delta\omega,T)}{S(q,T)} \quad (4)$$

where $S_{el}(q,\Delta\omega) = \int_{-\Delta\omega}^{\Delta\omega} d\omega S(q,\omega)$ is the elastic component (defined by the frequency cut-off $\Delta\omega$ associated with the width of the elastic peak) and the structure factor $S(\vec{q},T)$ is

$S(\vec{q},T) = \int_{-\infty}^{\infty} d\omega S(\vec{q},\omega,T)$. The DW factor, as defined in Eq. 4, is simply the elastic fraction of the total scattering. An alternative definition of the amorphous DW factor has been discussed by a number of authors [23,24] and is of the form

$$D_w(q,T) \propto F_s(q,T,t_p) = \frac{1}{N}\left\langle \sum_j^N \exp\left(i\vec{q}\cdot\Delta\vec{r}_j(t_p)\right) \right\rangle \quad (5)$$

where $F_s$ is the self intermediate scattering function evaluated at time $t_p$ corresponding to the plateau in the relaxation function. i.e. the non-ergodicity parameter [23]. The average in Eq. 5 is over different initial configurations. While different in form to Eq. 4, Eq. 5 also defines the DW factor as the fraction of scattering associated with elastic behavior, except the separation between the two types of motion is made using time rather than frequency.

The difference between the definitions of the DW factor for crystals and amorphous solids is striking. The amorphous definition requires that the inelastic contribution is measured and, in doing so, entangles the definition of the DW factor with the relaxation dynamics. In contrast, the DW factor of the crystal is defined in terms of the static scattering only and, as a result, is clearly identified as a purely structural measure. Can we obtain the amorphous DW factor using only static scattering data, just as in the crystal case? This capability would provide a number of significant benefits. First, it would allow us to use X-ray scattering, our main tool for structure determination in materials, to obtain DW factors. Secondly, it avoids the unnecessary need to resolve the frequency or time dependence of the scattering intensities. Finally, it would clarify the structural nature of the DW factor and, in so doing, support the interpretation of the DW factor as an order parameter for amorphous solidification. We shall now proceed to demonstrate that an analysis of the temperature dependence of elastic scattering from a coherent radiation source can indeed extract the DW factor for an amorphous material.

**3. Amorphous Scattering With and Without Angular Averaging**

The total structure factor $S(\vec{q})$ is obtained from the measured scattering intensity $I(\vec{q})$ at the wavevector $\vec{q}$ through the relation [25]

$$S(\vec{q}) = 1 + \frac{I(\vec{q}) - \sum_{i=1}^{N} f_i^2(q)}{\frac{1}{N}\left[\sum_{i=1}^{N} f_i(q)\right]^2}$$

$$= 1 + \frac{\left\langle \sum_{j \neq k}^{N} f_j f_k \exp\left(-i\vec{q} \cdot (\vec{r}_j - \vec{r}_k)\right) \right\rangle_\tau}{\frac{1}{N}\left[\sum_{i=1}^{N} f_i(q)\right]^2}$$

(6)

where $f_j$ and $\vec{r}_j$ are the form factor and position of atom j, N is the number of atoms and $\langle \ldots \rangle_\tau$ is the average of the particle positions taken over some measurement time τ. In selecting the measurement time τ one would ideally want the minimum value large enough to satisfy two requirements. The first is that enough photons are collected to provide an accurate determination of the scattering probability at each wavevector. The operational test for this condition is that there should be little variation in the measured intensity between two sequential measurements from a sample well below its glass transition temperature. The second requirement is that the measurement time be of sufficient duration to properly average over vibrational motion. In the case of computer simulations, we only need consider the second requirement since we calculate the scattering probability directly. In experiments, this situation is generally reversed. With Debye frequencies of the order of $10^{12}$ Hz, vibrational averaging is achieved within an extremely short time, no more than $10^{-6}$ s. Experimentally, the measurement time is entirely determined by the requirement to accumulate a sufficient number of photons per pixel to satisfy our operational test [27].

The scattering intensity $S(\vec{q},T)$ defined in Eq. 6 retains the orientational dependence of the scattering vector $\vec{q}$. The structure factor $\bar{S}(q,T)$ reported for liquids and glasses corresponds to the orientational average $\bar{S}(q,T) = \langle S(\vec{q},T) \rangle_{\hat{q}}$ where $\langle ... \rangle_{\hat{q}}$ is the average over the orientation of the wavevector at a fixed magnitude $q$ and temperature T.

In this paper we have calculated $S(\vec{q},T)$ for the binary $A_{80}B_{20}$ model glass former introduced by Kob and Anderson [28]. The molecular dynamics simulations are performed at fixed NVT with the LAMMPS software [29], using a Nose-Hoover thermostat to keep the temperature constant. Our system consists of a binary mixture of 5,000 particles at a density of 1.20. All quantities are reported in reduced Lennard-Jones units as described in Ref [28]. The time step is set to 0.002 and a thermostat damping parameter of 0.2 is used. The simulations are initiated in relaxed glassy configurations generated using the Swap Monte Carlo algorithm at T=0.05 as described in Ref [13]. The energy of these configurations is then minimized using the LAMMPS implementation of the Polak-Ribiere conjugate gradient algorithm to give the T=0 structure. For each temperature considered, the system is initialized in the T=0 configuration, the thermostat is then set to the reported temperature, and the system is equilibrated for $10^6$ timesteps before measurements are taken. Ten repeats of the simulation are performed starting from different initial configurations. The error bars correspond to the standard error, which are omitted when they are of comparable size to the symbol. The structure factor is calculated from the simulated configurations directly using Eq. 6. The measurement time is set to $\tau = 2 \times 10^4$ and 1,000 configurations are evenly sampled over this time interval and used to compute the time average. In calculations of the scattered intensities we have set the form factors for species A and B to be equal.

In Fig. 1a we plot the structure factor $\bar{S}(q,T)$ as a function of q for a range of temperatures. The value of the wavevector at the nth peak is designated $q_n$ so that the large first peak corresponds to q = $q_1$. We note that the modest dependence on T already well established in the literature [30]. 'Modest', we acknowledge, does not necessarily mean without interest. Mauro et al [31] reported a difference between the value of $\bar{S}(q_1,T)$ at the glass transition temperature and a value estimated by an extrapolation from higher T that appears to correlate with the fragility. We do see, in Fig. 1b, evidence of a change in the temperature derivative of $\bar{S}(q_1,T)/\bar{S}(q_1,0)$, similar to that seen experimentally [32,33].

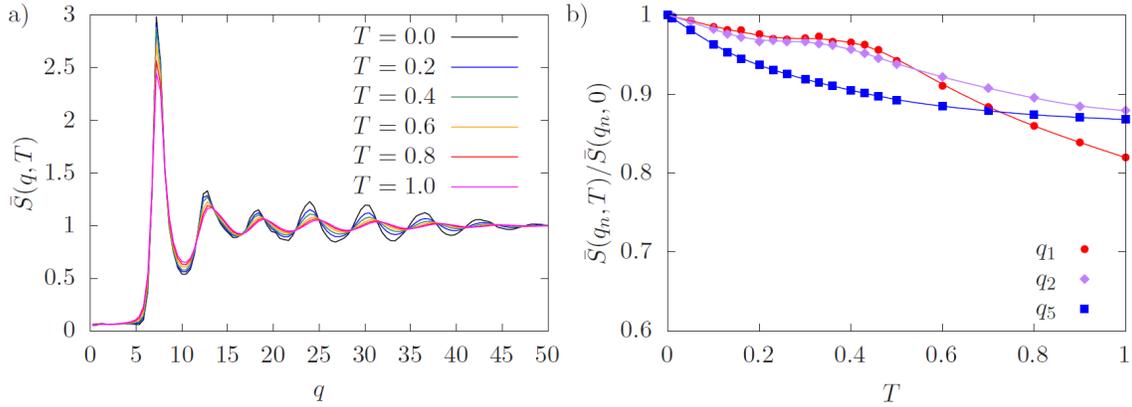

**Figure 1.** a) The dependence of the structure factor $\bar{S}(q,T)$ on the wavevector q for the simulated binary mixture for a range of temperatures. T and q are reported in reduced Lennard-Jones units as described in Ref. [28]. b) The temperature dependence of $\bar{S}(q_n,T)/\bar{S}(q_n,0)$ (for $q_n = q_1$, $q_2$ and $q_5$ where $q_n$ is the position of the nth peak in $\bar{S}(q,T)$.

Even though a transformation between fluid and solid has taken place over the temperature range studied, one accompanied by a large change in the amplitude of particle displacements, the weak temperature dependence of $\bar{S}(q,T)$ provides little indication of this. The culprit, it turns out, is the orientational averaging of the intensity. The angle-dependent scattering $S(\vec{q},T)$ from an amorphous solid exhibits a complex granularity over the wavevector space referred to as *speckle* [33]. This speckle, as shown in Fig. 2, disappears on heating, leaving a uniform scattering intensity over of each sphere in q-space characterized by a fixed wavevector magnitude q. The temperature dependence of this smoothing process is erased when we average over the wavevector orientation $\hat{\vec{q}}$. The problem we address in this paper is to extract a useful measure of structural restraint from the temperature dependence of $S(\vec{q},T)$ *before* any averaging over wavevectors.

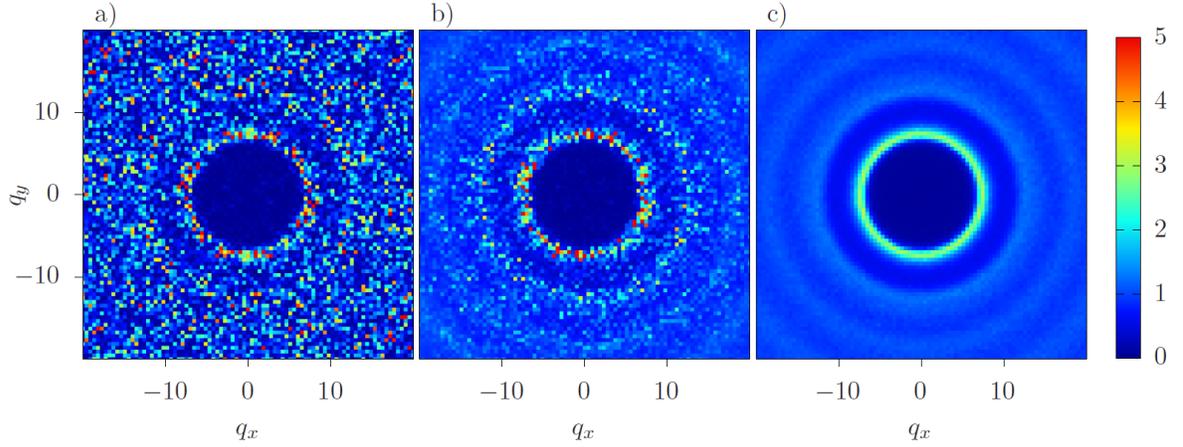

**Figure 2.** The scattering intensity $S(\vec{q})$, with intensities indicated by color as shown, in the ($q_x$,$q_y$) plane for a) T = 0.0 b) T = 0.3 and c) T = 0.6. The reduction of the intensity fluctuations (speckle) with increasing temperature is evident.

## 4. Analysis of the Temperature Dependence of the Speckle Pattern

We shall consider the explicit separation of the thermal fluctuations from the static disorder of the parent glass. Let $\vec{r}_j = \vec{R}_j + \vec{u}_j$ where $\vec{R}_j$ is the position of the atom j in the parent T=0 configuration and $\vec{u}_j$ its thermally excited displacement. Then

$$S(\vec{q}) = 1 + \frac{1}{N} \sum_{j \neq k} \exp\left[-i\vec{q} \cdot (\vec{R}_j - \vec{R}_k)\right] \left\langle \exp(-iP_{jk}(\vec{q})) \right\rangle_\tau \qquad (7)$$

where $P_{jk}(\vec{q}) = \vec{q} \cdot (\vec{u}_j - \vec{u}_k)$. If the distribution of thermal displacements sampled during the measurement time τ can be treated as Gaussian [34], we can write

$$\left\langle \exp(-iP_{ij}(\vec{q})) \right\rangle_\tau = \exp\left(-<P_{ij}^2(\vec{q})>_\tau /2\right) \qquad (8)$$

Next, we shall separate this fluctuation term into its spatial mean and the local fluctuations about this mean, i.e.

$$\exp(-<P_{jk}^2(\vec{q})>_\tau /2) = <\exp(-<P_{jk}^2(\vec{q})>_\tau /2)>_{jk} + h_{jk}(\vec{q}) \qquad (9)$$

so that

$$S(\vec{q},T)-1 = <\exp(-<P_{jk}^2(\vec{q})>_\tau /2)>_{jk} [S(\vec{q},0)-1] + \left(\frac{1}{N}\sum_{j\neq k}\exp(-\vec{q}\cdot\vec{R}_{jk})h_{jk}(\vec{q})\right) \qquad (10)$$

This derivation has, so far, followed the standard treatment for the Debye-Waller factor for a crystal. The next step for the crystal would be to select only the wavevectors corresponding to peaks in intensity and determine the values of $S(\vec{q},0)$ from the analytic expressions for the perfect crystal. Amorphous scattering (like diffuse scattering from disordered crystals) distributes significant intensity across the continuum of $\vec{q}$ space so that we need to generalize the analysis of the thermal contribution to disorder.

We propose the following. We shall partition phase space in terms of the magnitude of $S(\vec{q},0)-1$ so that all of the values of $\hat{\vec{q}}$ (for a given magnitude q) for which $S(\vec{q},0)-1$ lies within some narrow range about a selected value are grouped together. Then we can write

$$<S(\vec{q},T)-1>_{S_o} = <<\exp(-<P_{jk}^2(\vec{q})>_\tau /2)>_{jk}>_{S_o} [S(\vec{q},0)-1] + <\left(\frac{1}{N}\sum_{j\neq k}\exp(-\vec{q}\cdot\vec{R}_{jk})h_{jk}(\vec{q})\right)>_{S_o}$$

$$(11)$$

where $<...>_{S_o}$ indicates a restricted average of that set of wavevectors associated with a given value of $S(\vec{q},0)-1$. Next, we shall assume that the two restricted averages on the RHS of Eq. 11 are independent of the value of $S(\vec{q},0)-1$ so that we can rewrite Eq. 11 as

$$<S(\vec{q},T)-1>_{S_o} = D_w(q,T)[S(\vec{q},0)-1] + H(q,T) \qquad (12)$$

Eq.12 predicts a linear relation between the intensity at some non-zero T and the T=0 value with a slope and intercept that depend on T and q. As shown in Fig. 3, this prediction is empirically confirmed over a range of temperatures. This confirmation allows us to unambiguously extract the values of the $D_w(q,T)$ and $H(q,T)$ from scattering data and is the major result of this paper. $D_w(q,T)$ quantifies the residual correlation between the scattering at a non-zero T and T=0; so that $D_w = 1$ at T = 0 and goes to zero at the temperature at which the scattering is uniformly distributed (as shown in Fig. 2). While the use of the T = 0 scattering represents the ideal reference, we have demonstrated that a similar linear relation is obtained for a non-zero reference T well below $T_g$ (see Supplementary Materials).

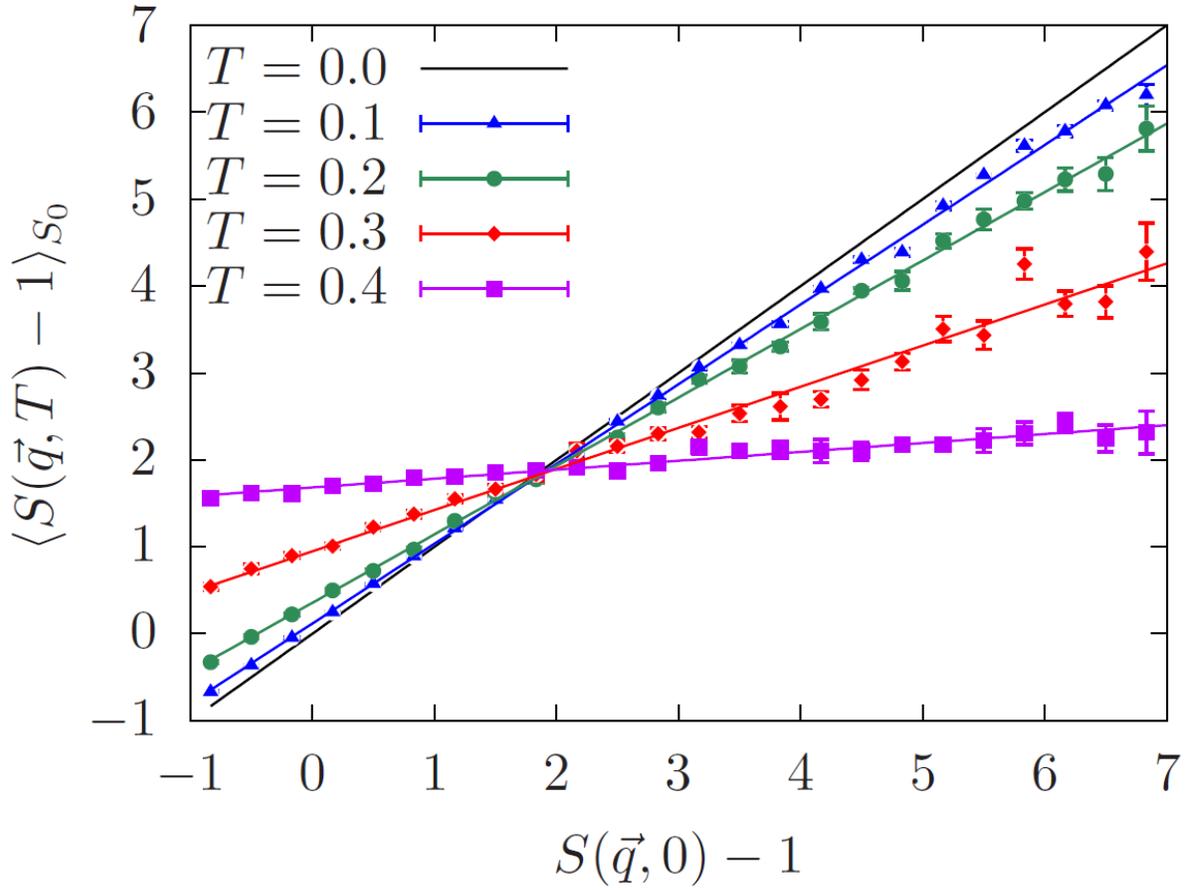

**Figure 3.** The variation of $\langle S(\vec{q},T)\rangle_{S_o}$ with respect to the selected value of $S(\vec{q},0)$ (see text). Each color corresponds to a different temperature as indicated. The straight lines are least square fits.

The dramatic decrease of $D_w(q_1,T)$ from 1 to 0 on heating is in sharp contrast to the weak dependence of the structure factor itself, as shown in Fig. 4. As $D_w(q_1,T)$ is obtained from an analysis of the elastic scattering intensities, it represents a measure of structure that differentiates the liquid and the amorphous solid and, hence, satisfies the requirements of an experimentally accessible structural order parameter for amorphous solidification. This was the goal of our study.

We can confirm that $D_w(q,T)$, as obtained from the linear fits in Fig. 3, is indeed the DW factor by comparing it to the theoretical value $D_o(q,T)$ based on a harmonic expansion [25],

$$D_o(q,T) \equiv \exp\left(-\frac{q^2}{3}\left\langle <u_j^2>_\tau \right\rangle_j\right) \qquad (13)$$

where we have neglected any correlation between the motions of different particles. As demonstrated in Fig. 4, $D_w(q_1,T) \approx D_o(q_1,T)$. Further confirmation of this connection is provided by the dependence of $\ln[D_w(q_1,T)]$ on $q_1^2$ (see Supplementary Material) where we find, at low T, the linear relation predicted by Eq. 13.

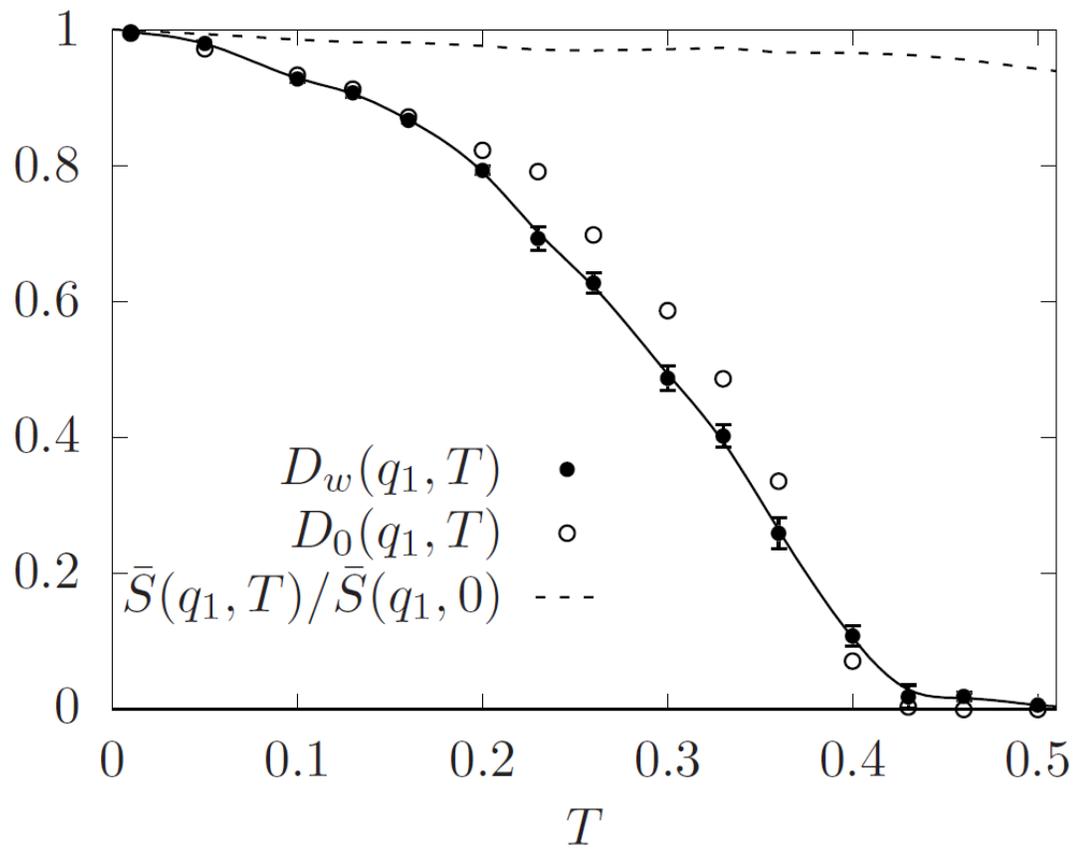

**Figure 4.** The temperature dependence of $D_w(q_1,T)$ and, for comparison, that of $\bar{S}(q_1,T)/\bar{S}(q_1,0)$. Also plotted is $D_o(q_1,T)$ as defined in Eq.13. The curve is fitted to $D_w(q_1,T)$ as a guide to the eye.

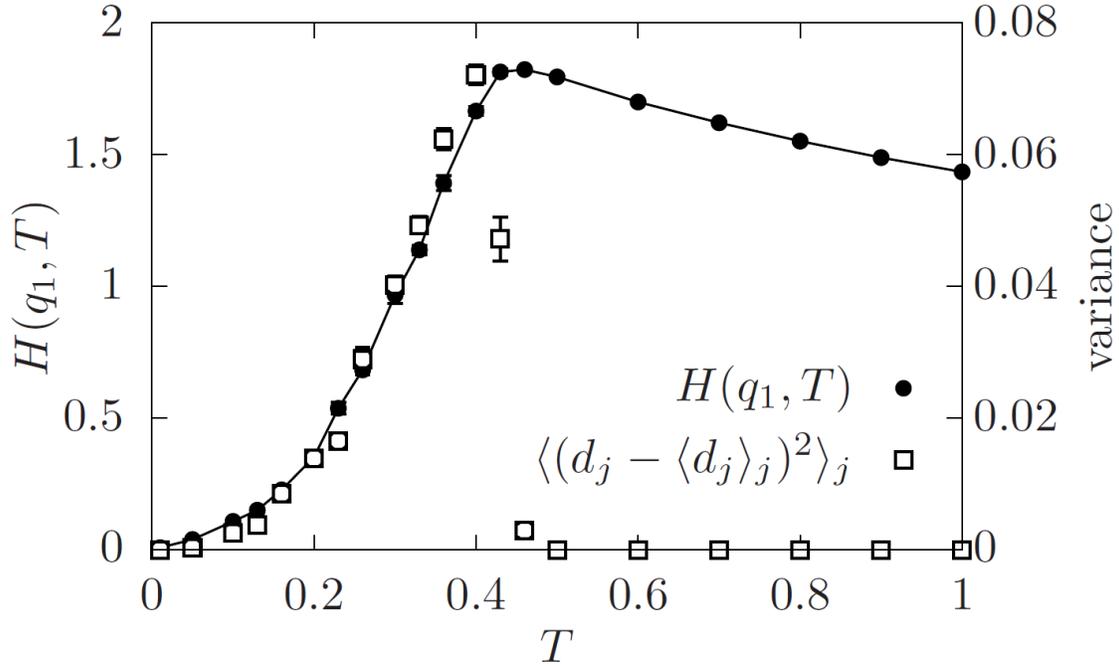

**Figure 5.** The quantity $H(q_1,T)$, obtained as the intercept from the analysis in Fig. 3, as a function of T. Plotted on the right axis is the variance $\left\langle \left(d_j - <d_j>_j\right)^2 \right\rangle_j$ where $d_j = \exp\left(-\frac{q_1^2}{3}<u_j^2>_\tau\right)$. The curve is fitted to $H(q_1,T)$ as a guide to the eye.

The plot in Fig. 3 also allows us to extract $H(q_1,T)$ from the intercept, which we have plotted against temperature in Fig. 5. As indicated in the derivation of Eq. 11, $H(q_1,T)$ is associated with the spatial variation in the amplitude of the thermal fluctuations in atomic positions. To confirm this connection, we have calculated the spatial variance in the quantity $d_j = \exp\left(-\frac{q_1^2}{3}<u_j^2>_\tau\right)$ and plotted the results in Fig. 5. Apart from a scaling constant, the variance closely matches the values of $H(q_1,T)$ up to value of T at which the variance abruptly vanishes. The dramatic rise and fall in $\left\langle\left(d_j-<d_j>_j\right)^2\right\rangle_j$ reflects the twin effects of increasing temperature: enhanced fluctuations and increased relaxation rate. The former produces the increase in the variance, observed at low T, while the increasing rate of relaxation ends up smoothing out fluctuations (as shown in Fig. 2). The competing effects of increased amplitude of thermal fluctuations and increased relaxation rate of these fluctuations with increasing T results in a crossover in $H(q_1,T)$; at lower values of T, H provides a measure of the variance of the structural restraint while above this crossover temperature $H(q_1,T) = \bar{S}(q_1,T)$, the structure factor. The existence of the maximum appears to be universal, but its exact position is dependent on the choice of measurement time (see Supplementary Materials). These results establish that our analysis of the speckle intensities can not only extract information about the mean value of the atomic restraint but information about its spatial heterogeneity, in the form of the variance, as well. This is an exciting result as there is a growing theoretical consensus [35-38] that the mechanical properties of glasses are best understood as a consequence of a spatial distribution of rigidity. On the length scales of individual atoms, this mechanical heterogeneity is manifest as a distribution of atomic restraint [39] which we have demonstrated can be measured (to within a temperature independent constant).

## 5. Conclusions

In conclusion, we have established, in principle, a new method to directly measure the amorphous Debye-Waller factor from the static scattering intensities from a coherent radiation source. The analysis presented here returns not only the DW but a new quantity, labeled H(q,T) in Eq. 12, which we have shown to be proportional to the variance in the atomic displacements and, hence, represents a direct experimental measure of the heterogeneity of atomic restraint. Unlike approaches based on dynamic measurements, the measurement of $D_w(q_1,T)$ we present follows closely the measurement of the analogous quantity in crystals and so represents a general methodology for monitoring the solidification of a material irrespective of whether the solid state is ordered or disordered. As our approach does not require time resolved intensities, the measurement time at each temperature can be short, less than a second in the case of synchrotron radiation [40], a considerable advantage when contending with radiation-induced damage of the sample [40].

How feasible is the experiment we have modelled in this paper ? Speckle intensities are routinely resolved and measured using X-ray photon correlation spectroscopy for metallic [41], oxide [40] and colloidal glasses [42] over a wide range of temperatures. These experiments make use of coherent X-rays generated at synchrotron facilities. As the measurements proposed here require exactly the same measured intensities, there appears to be no fundamental obstacle to experimentally testing of the linearity reported in Fig. 3 and the extraction of the Debye-Waller factor $D_w(q_1,T)$ and the associated quantity $H(q_1,T)$. Simulation results indicate that the analysis we have presented applies even when thermal expansion is allowed at constant pressure (see Supplementary Materials).

Previously [12,13], we have argued that structural restraint provides a much more useful characterization of amorphous structure than more traditional measures based on topology or geometry. The DW factor represents an experimentally accessible analogue of this measure of atomic restraint. In the preceding discussion we have made use of an operational classification of the adjective 'structural', one based on the static character of the experiment. Here we acknowledge that this definition, which we believe is useful, is not without ambiguity. A technique called Speckle-Visibility Spectroscopy (SVS) [43] uses the dependence of the resolution of a speckle patten on the magnitude of the measurement time τ to extract the relaxation functions, i.e. a dynamic experiment. In contrast, we propose using a fixed value of τ (and, hence, a static measurement) but to do so for a range of temperatures. The observed decrease in the correlation of the intensity at a given wavevector with increasing T, which we extract in the form of $D_w(q_1,T)$, includes a contribution due to the decrease in the structural relaxation time relative to the fixed τ, the same phenomenon that serves as the basis for SVS. The unavoidable difficulty is that in amorphous materials the influence of thermally excited vibrations is entangled with the influence of thermally activated relaxation. We cannot include one without the other. If we were apply a more stringent definition of 'structural' that excluded any property that exhibited a dependence on the measurement time, we would only be able to call the low and high temperature values of $D_w(q_1,T)$ structural, leaving a blank in the intermediate temperatures over which the transformation from liquid to amorphous solid actually takes place. The justification for our less stringent definition of a structural property, i.e. a property obtained from a static scattering measurement, lies in being able to demonstrate its ultimate usefulness in providing a framework for analyzing glassy phenomenology, a usefulness that, at this preliminary stage, we can only argue for based on the success on the analogous structural restraint parameter introduced in recent simulations [12,13].

We emphasize that irrespective of whether one accepts the structural interpretation or opts to see the DW factor as a dynamic property, the value of $D_w(q_1,T)$ and $H(q_1,T)$ as an accessible means of characterizing the capacity of amorphous configurations to restrain atomic motion is not in question. A number of groups have reported abrupt changes in the temperature derivative of either the structure factor [25,31] or the radial distribution function [44,45] in deeply supercooled liquids. It will be interesting to see if there are any correlations between these observations and the temperature dependence of $D_w(q_1,T)$. In elevating the Debye-Waller factors to a fundamental characteristic of structure, we are presented with the opportunity to rethink the very framework of glassy phenomenology; to explore how the temperature dependence of fundamental features of amorphous materials, such as fragility, elastic response, yield stress and thermal conductivity, are related directly to the measured change in the degree of particle restraint.

**Supplementary Material**

The Supplementary Material document provides additional information on the following topics:

1. The q Dependence of $D_w(q,T)$

2. The Comparison of the Debye-Waller factor from Elastic Scattering and that based on the Self-Intermediate Scattering Function $F_s(q,t)$

3. The Dependence of H(q,T) on the Measurement Time τ

4. The Linear Relation between [S(q,T)-1] and [S(q,T$_{ref}$)-1] for T$_{ref}$ >0 and Under Constant Pressure

**Acknowledgements**

CFP and PH gratefully acknowledge Gang Sun for providing equilibrated initial atomic configurations and helpful discussions with Amelia Liu and Tim Petersen. CFP is the recipient of an Australian Research Council Discovery Early Career Award (DE210100256) funded by the Australian Government. This research was undertaken with the assistance of resources and services from the National Computational Infrastructure (NCI), which is supported by the Australian Government.

**Author contributions.** Conceptulization - PH; Methodology – CFP + PH; Calculations - CFP; Writing – CFP + PH

**Author Contact Information.** charlotte.petersen@unimelb.edu.au & peter.harrowell@sydney.edu.au

# References

1. Definition of the glass transition from IUPAC, *Compendium of Chemical Terminology*, 2nd Ed. (1997).

2. P. Steinhardt, D. Nelson, and M. Ronchetti, Bond orientational order in liquids and glasses. *Phys. Rev. B* **28**, 784 (1983).

3. W. Lechner and C. Dellago, Accurate determination of crystal structures based on averaged local bond order parameters. *J. Chem. Phys.* **129**, 114707 (2008).

4. Y. Hiwatari, T. Saito and A. Ueda, Structural characterization of soft-core and hard-core glasses by Delaunay tessellation. *J. Chem. Phys.* **81**, 6044 (1984).

5. J. D. Honeycutt and H. C. Andersen, Molecular dynamics study of melting and freezing of small Lennard-Jones clusters. *J. Phys. Chem*. **91,** 4950-4963 (1987).

6. S. Marin-Aguilar, H. H. Wensink, G. Foffi and F. Smallenburg, Tetrahedrality dictates dynamics in hard sphere mixtures. *Phys. Rev. Lett*. **124**, 208005 (2020).

7. K. J. Laws, D. B. Miracle and M. Ferry, A predictive structural model for bulk metallic glasses. *Nature Comm*. **6**, 8123 (2015).

8. C. P. Royall and S. R Williams, The role of local structure in dynamical arrest. *Phys. Rep.* **560**, 1-75 (2015).

9. C. De Michele, S. Gabrielli, P. Tartaglia, F. Sciortino, Dynamics in the presence of attractive patchy interactions. *J. Phys. Chem. B* **110**, 8064–8079 (2006).

10. S. Marin-Aguilar, H. H. Wensink, G. Foffi and F. Smallenburg, Slowing down supercooled liquids by manipulating their local structure. *Soft Matter* **15**, 9886-9893 (2019).

11. D. Wei, J. Yang, M.-Q. Jiang, L.-H. Dai, Y.-J. Wang, Dyre, I. Douglass and P. Harrowell, Assessing the utility of structure in amorphous materials. *J. Chem. Phys.* **150**, 114502 (2019).

12. G. Sun, L. Li and P. Harrowell, The structural difference between strong and fragile liquids. *J. Non-Cryst. Sol*. **13**, 100080 (2022).

13. G. Sun and P. Harrowell, A general structural order parameter for the amorphous solidification of a supercooled liquid. *J. Chem. Phys*. **157**, 024501 (2022).

14. G. Sun and P. Harrowell, Amorphous solidification of a supercooled liquid in the limit of rapid cooling. *J. Chem. Phys.* **158**, 204506 (2023)

15. G. Sun, S. Saw, I. Douglass and P. Harrowell, Structural origin of enhanced dynamics at the surface of a glassy alloy. *Phys. Rev. Lett*. **119**, 245501 (2017).

16. P. Tarazona, A density functional theory of melting. *Mol. Phys*. **52**, 81 (1984).87

17. P. Debye, *Annalen der Phys*. **348**, 49-92 (1913).

18. J. Als-Nielsen and D. McMorrow, *Elements of Modern X-ray Physics*. (John Wiley, New York, 2011).

19. D. W. J. Cruickshank, The determination of the anisotropic thermal motion of atoms in crystals. *Acta Cryst*. **9**, 747 (1956).

20. *The Rietveld Method*, ed. R. A. Young (Oxford University Press, London, 1993).

21. B. Frick, D. Richter, W. Petry and U. Buchenau, Study of the glass transition order parameter in amorphous polybutadiene by incoherent neutron scattering. *Z. Phys. B-Cond. Matt.* **79**, 73-79 (1988).

22. U. Buchenau and R. Zorn, A relation between fast and slow motions in glassy and liquid Selenium. *EPL* **18**, 523 (1992).

23. F. Sciortino and W. Kob, The Debye-Waller factor of liquid silica: Theory and simulation. *Phys. Rev. Lett*. **86**, 648-651 (2001).

24. S. Jabbari-Farouji, H. Tanaka, G. H. Wegdam and D. Bonn, Multiple nonergodic disordered state in Laponite suspensions: A phase diagram. *Phys. Rev. E* **78** 061405 (2008).

25. N. A. Mauro, A. J. Vogt, M. L. Johnson, J. C. Bendert, R. Soklaski, L. Yang and K. F. Kelton, Anomalous structural evolution and liquid fragility signatures in Cu-Zr and Cu-Hf liquids and glasses, *Acta Mater.* **61**, 7411-7421 (2013).

26. B. E. Warren, *X-ray Diffraction* (Dover, London, 1969).

27. N. A. Dyson. *X-Rays in Atomic and Nuclear Physics* (2$^{nd}$ Edition) (Cambridge UP, Cambridge, 1990).

28. W. Kob and H. C. Andersen, Testing mode-coupling theory for a supercooled binary Lennard-Jones mixture I: The van Hove correlation function. *Phys. Rev. E* **51**, 4626-4641 (1995).

29. A. P. Thompson et al., LAMMPS - a flexible simulation tool for particle-based materials modeling at the atomic, meso, and continuum scales, *Comp Phys Comm*, **271**, 10817 (2022).

30. E. Bartsch, H. Bertagnolli, P. Chieux, A. David and H. Sillescu, Temperature dependence of the static structure factor of ortho-terphenyl in the supercooled liquid regime close to the glass transition. *Chem. Phys*. **169**, 373-378 (1993).

31. N. A. Mauro, M. Blodgett, M. L. Johnson, A. J. Vogt and K. F. Kelton, A structural signature of liquid fragility. *Nature Com*. **10**, 1038 (2014).

32. C. W. Ryu, W. Dmowski, K. F. Kelton, G. W. Lee, E. S. Park, J. R. Morris and T. Egami, Curie-Weiss behavior of liquid structure and ideal glass state. *Sci. Rep*. **9**, 18579 (2018).

33. J. W. Goodman, Some fundamental properties of speckle. *J. Opt. Soc. Am*. **66**, 1145-1150 (1976).

34. L. V. Meisel and P. J. Cote, Structure factors in amorphous and disordered harmonic Debye solids. *Phys. Rev. B* **16**, 2978-2980 (1977).

35. M. Tsamados, A. Tanguy, C. Goldenberg and J.-L. Barrat, Local elasticity map and plasticity in a model Lennard-Jones glass. *Phys. Rev. E* **80**, 026112 (2009).

36. H. Wagner, D. Berdorf, S. Kuchemann, M. Schwabe, B. Zhang, W. Arnold and K. Samwer, Local elastic properties of a metallic glass. *Nature Mat*. **10**, 439-442 (2011).

37. G. Kapteijins, D. Richard, E. Bouchbinder and E. Lerner, Elastic moduli fluctuations predict wave attenuation rate in glasses. *J. Chem. Phys*. **154**, 081101 (2021).

38. M. Lerbinger, A. Barbot, D. Vandembroucq and S. Patinet, Relevance of shear transformations in the relaxation of supercooled liquids. *Phys. Rev. Lett*. **129**, 195501 (2022).

39. S. Saw and P. Harrowell, Rigidity in condensed matter and its origin in configurational constraint. *Phys. Rev. Lett*. **116**, 137801 (2016).

40. E. Alfinelli, F. Caporaletti, F. Dallari, A. Martinelli, G. Monaco, B. Ruta, M. Sprung, M. Zanatta and G. Baldi, Amorphous-amorphous transformation induced in glasses by intense X-ray beams. *Phys. Rev. B* **107**, 054202 (2023)

41. N. Neuber, et al, Disentangling structural and kinetic components of the α-relaxation in supercooled metallic liquids, *Comm. Phys*. **5**, 316 (2022).

42. F. A. de Melo Marques, R. Angelini, E. Zaccarelli, B. Farago, B. Ruta, G. Ruocco and B. Ruzicka, Structural and microscopic relaxations in a colloid glass. *Soft Matter* **11**, 466 (2015).

43. R. Bandyopadhyay, A. S. Gittings, S. S. Suh, P. K. Dixon and D. J. Durian, Speckle-visibility spectroscopy: A tool to study time-varying dynamics. *Rev. Sci. Instrum*. **76**, 093110 (2005).

44. F. Abraham, An isothermal–isobaric computer simulation of the supercooled-liquid/glass transition region: Is the short-range order in the amorphous solid fcc? *J. Chem. Phys*. **72,** 359 (1980).

45. D. Li, H. Xu and J. P. Witmer, Glass transition of two-dimensional 80-20 Kob-Andersen model at constant pressure. *J. Phys.: Cond. Matt.* **28**, 045101 (2016).

**SUPPLEMENTARY MATERIAL**

**Direct Measurement of the Structural Change Associated with Amorphous Solidification using Static Scattering of Coherent Radiation**

Charlotte F. Petersen[1,2] and Peter Harrowell[1,*]

[1]*School of Chemistry, University of Sydney, Sydney, New South Wales 2006 Australia*

[2]*School of Chemistry, University of Melbourne, Melbourne, Victoria 3010 Australia*

* Corresponding author: peter.harrowell@sydney.edu.au

Contents

*1. The q Dependence of $D_w(q,T)$*

*2. The Comparison of the Debye-Waller factor from Elastic Scattering and that based on the Self-Intermediate Scattering Function $F_s(q,t)$*

*3. The Dependence of H(q,T) on the Measurement Time τ*

*4. The Linear Relation between [S(q,T)-1] and [S(q,T_{ref})-1] for $T_{ref}$ >0 and Under Constant Pressure*

**1. The q Dependence of $D(q,T)$**

A characteristic of the Gaussian assumption used to derive the Debye-Waller factor, in general, and, specifically, in Eq. 13 is a dependence on q that goes as $\ln D_o(q,T) \propto q^2$. We can check to see whether $D_w(q,T)$ also exhibits this same dependence. In Fig. S1 we have plotted $\ln D_w(q,T)$ against $q^2$ and find the predicted linear relationship at low T, providing an additional support for our identification of $D_w(q,T)$ as the Debye-Waller factor. At higher temperatures, we do see signs of a deviation from this behaviour where the slope in Fig. S1 decreases with increasing q.

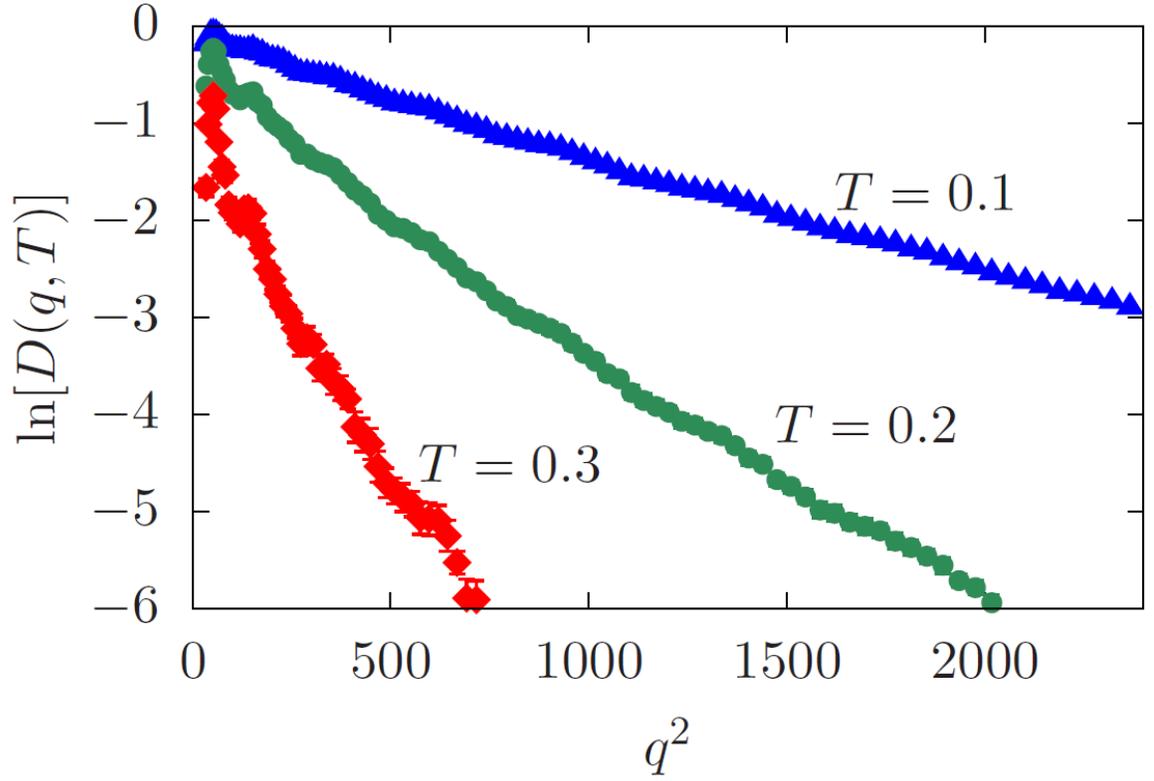

**Figure S1.** The dependence of $\ln D_w(q,T)$ on $q^2$ for three temperatures as indicated.

## 2. The Comparison of the Debye-Waller factor from Elastic Scattering and that based on the Self-Intermediate Scattering Function $F_s(q,t)$

The self-intermediate scattering function $F_s(q,t)$ is calculated using

$$F_s(q,t) = \frac{1}{N}\left\langle \sum_j^N \exp\left(i\vec{q}\cdot\Delta\vec{r}_j(t)\right)\right\rangle \tag{S2}$$

where $\Delta\vec{r}_j(t) = \vec{r}_j(t) - \vec{r}_j(0)$ and the average is taken over different initial configurations and different orientations of the wavevector $\vec{q}$ and plotted against log(t) in Fig. S2. An estimate of the Debye-Waller factor has been proposed [12,13], equal to the plateau height of $F_s(q,t)$. We shall define the plateau height by the measurement time $\tau = 2\times 10^4$ used in this present paper as indicated in Fig. S2.

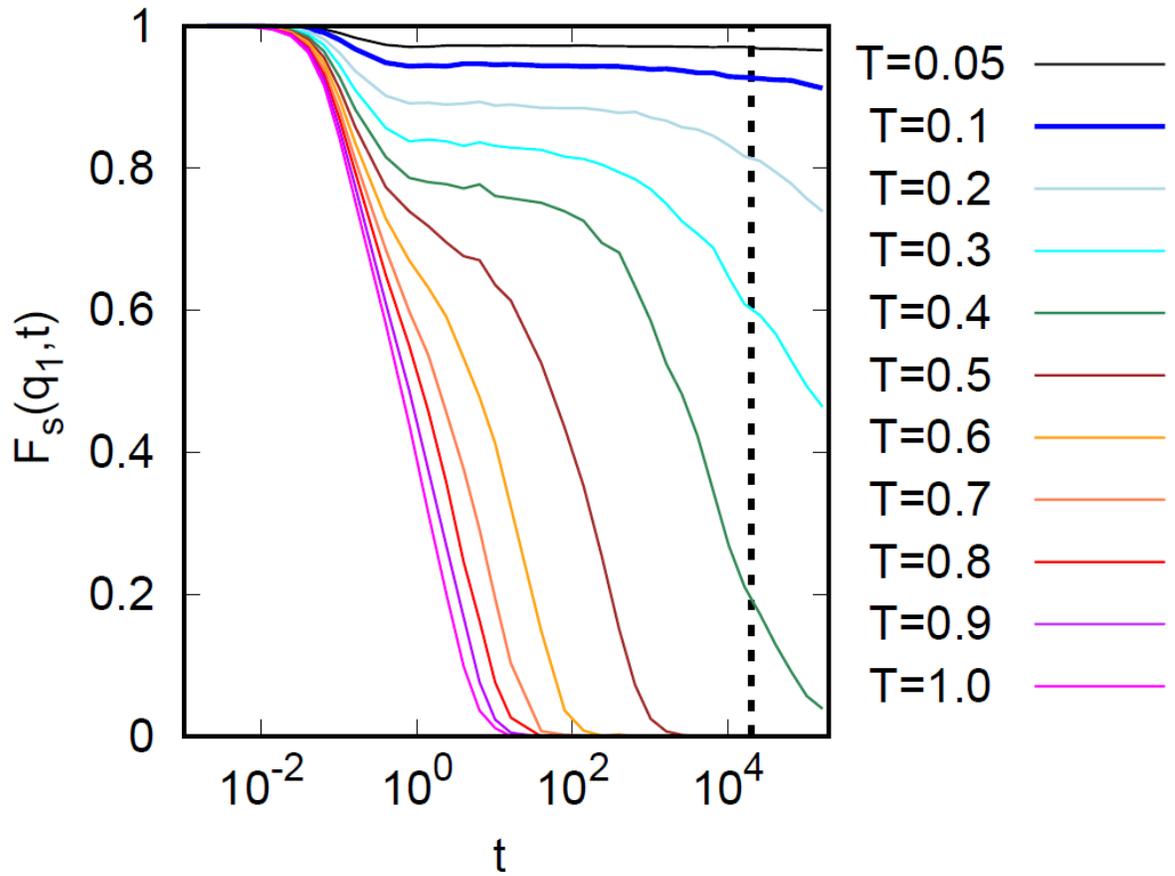

**Figure S2.** The self-intermediate scattering function $F_s(q,t)$ calculated at $q = q_1$ for a range of temperatures as indicated. The measurement time $\tau$ is indicated by the vertical dashed line.

In Fig. S3 we compare $D_o(q_1,T)$ with the DW factor $D_w(q_1,T)$ extracted from the scattering intensities and the quantity $D_o(q_1,T)$ obtained from the calculated mean squared displacements using the harmonic approximation. We find excellent agreement between $F_s(q,\tau)$ and $D_o(q_1,T)$. While there is some deviation between $D_o(q_1,T)$ and $D_w(q_1,T)$ at intermediate temperatures, there is strong agreement between the two quantities at both low T and the higher temperature at which both quantities vanish.

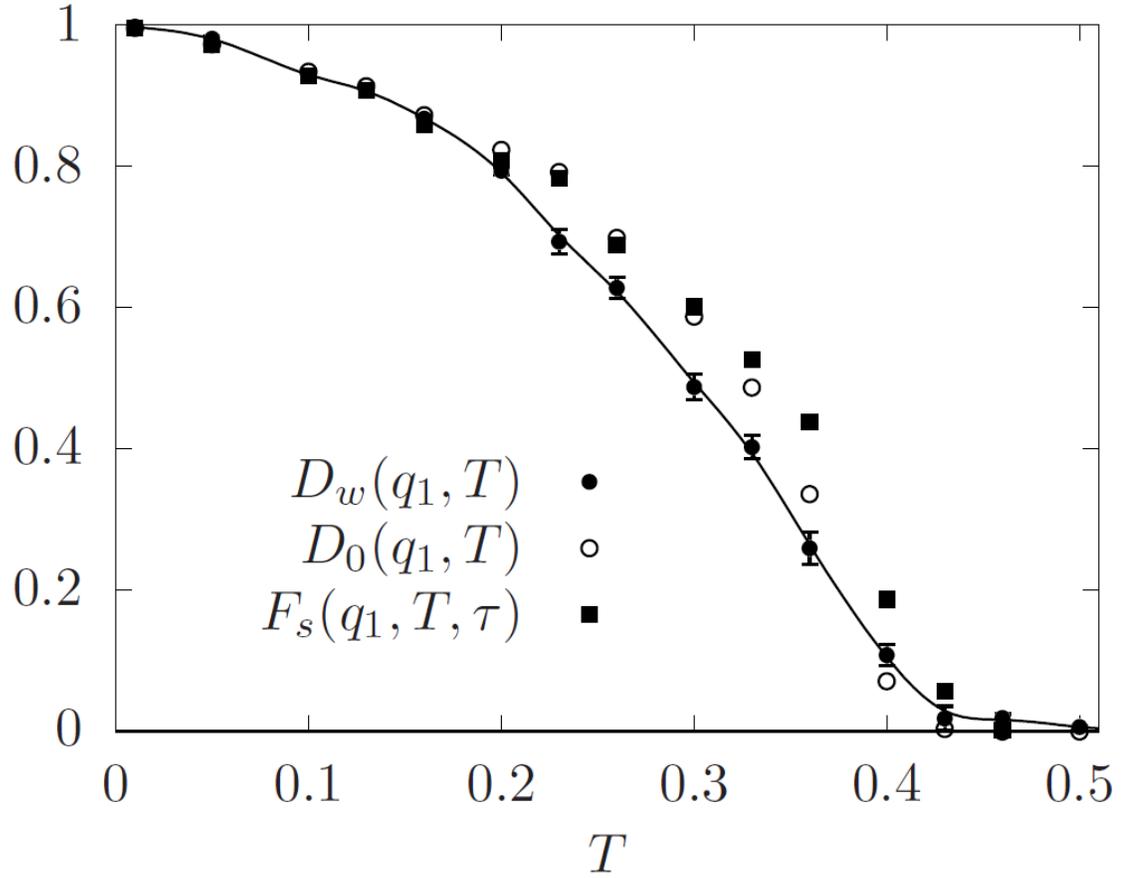

**Figure S3**. A comparison of the Debye-Waller factor $D_w(q_1,T)$ calculated from the elastic scattering and the harmonic approximation $D_o(q_1,T)$ with the estimate equal to $F_s(q_1,T,\tau)$. The time $\tau$ is the same measurement time as used in calculating $D_w(q_1,T)$.

### 3. The Dependence of H(q,T) on the Measurement Time $\tau$

In Fig. S4 we compare the temperature dependence of the quantity $H(q_1,T)$ and the variance of the local restraint $d_j = \exp\left(-\frac{q_1^2}{3}<u_j^2>_\tau\right)$ calculated using two different values of the measurement time, i.e. $\tau = 2 \times 10^4$ and $2 \times 10^6$. We find that the general shape of the curves including the cusps in both quantities are unaltered in form but that both maxima shift to a slightly lower temperature with the longer measurement time interval.

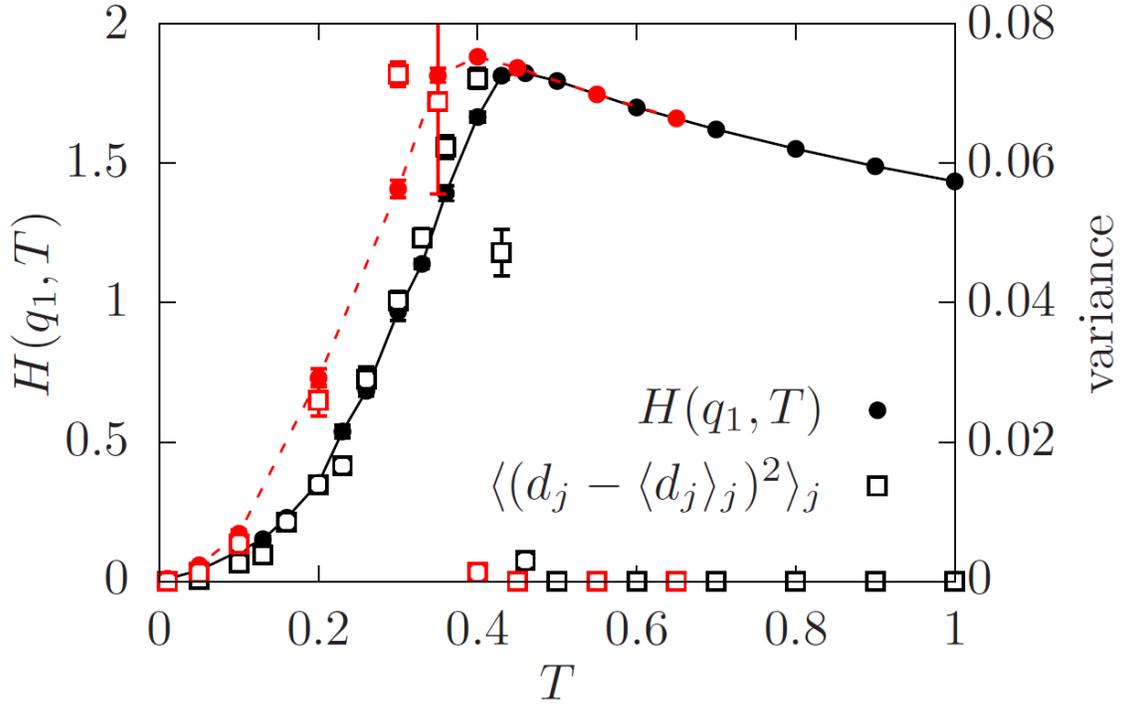

**Figure S4.** The quantity $H(q_1,T)$, obtained as the slope from the analysis in Fig. 3, as a function of T. Plotted on the right axis is the variance $\langle (d_j - \langle d_j \rangle_j)^2 \rangle_j$ where $d_j = \exp\left(-\frac{q_1^2}{3}\langle u_j^2 \rangle_\tau\right)$. Two different values of the measurement time have been used: $\tau$ = 2 x $10^4$ (black symbols) and 2 x $10^6$ (red symbols). The curve is fitted to $H(q_1,T)$ as a guide to the eye.

### 4. The Linear Relation between [S(q,T)-1] and [S(q,T$_{ref}$)-1] for T$_{ref}$ >0 and Under Constant Pressure

The linear dependence of <[S(q,T)-1]>$_{S(q,0)}$ on [S(q,0)-1], as plotted in Fig. 3, is a central result of the paper and crucial for extracting values of $D_w(q,T)$ and $H(q,T)$. In this Section we address the question of how persistent in this linear relation if a) we cannot access T = 0 for our reference configuration, and b) if the measurements are carried out at fixed pressure rather than fixed density so that expansion contributes to the decorrelation of speckle intensity during heating.

To address the first question, we have repeated the analysis of Fig.3 but for a reference temperature of T = 0.05 (as opposed to T = 0). As shown in Fig. S5 we recover similar linear dependence between the low T reference scattering and the higher T scattering intensities. The values for $D_w(q_1,T)$ and $H(q_1,T)$ obtained from the linear fits to the data in Fig. S5 is presented in Table S1 along with corresponding data obtained using $T_{\text{ref}} = 0$.

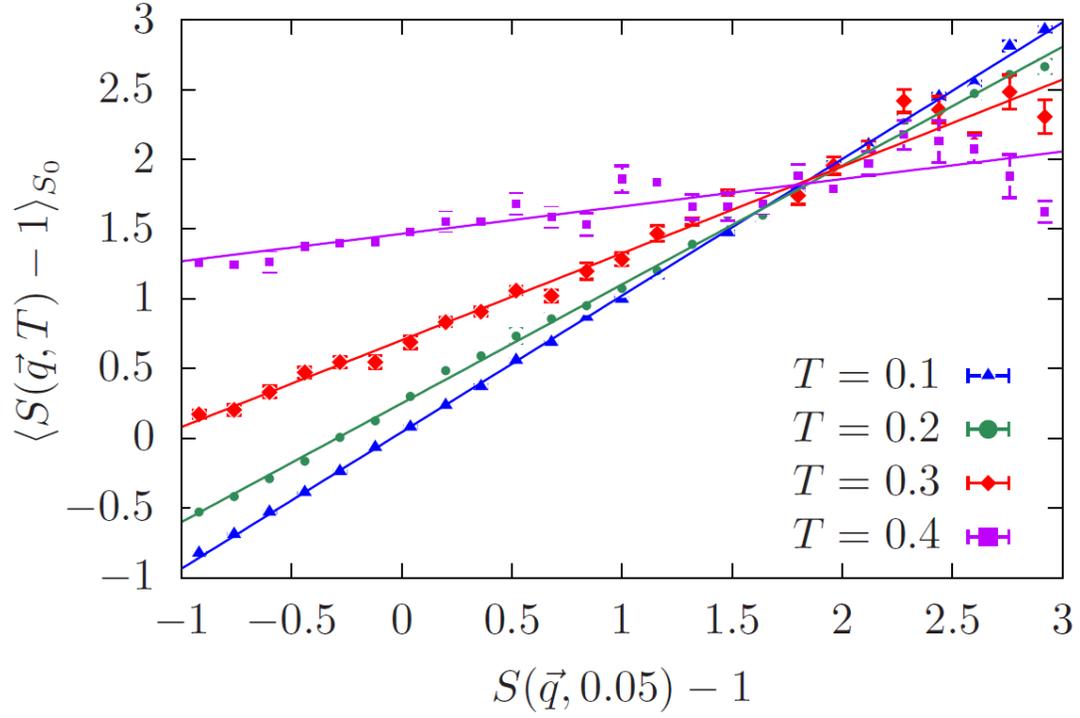

**Figure S5**. Plot of $\langle S(\vec{q},T)-1 \rangle_{S_o}$ vs $S(\vec{q},0.05)-1$ at constant density for 4 temperatures as indicated.

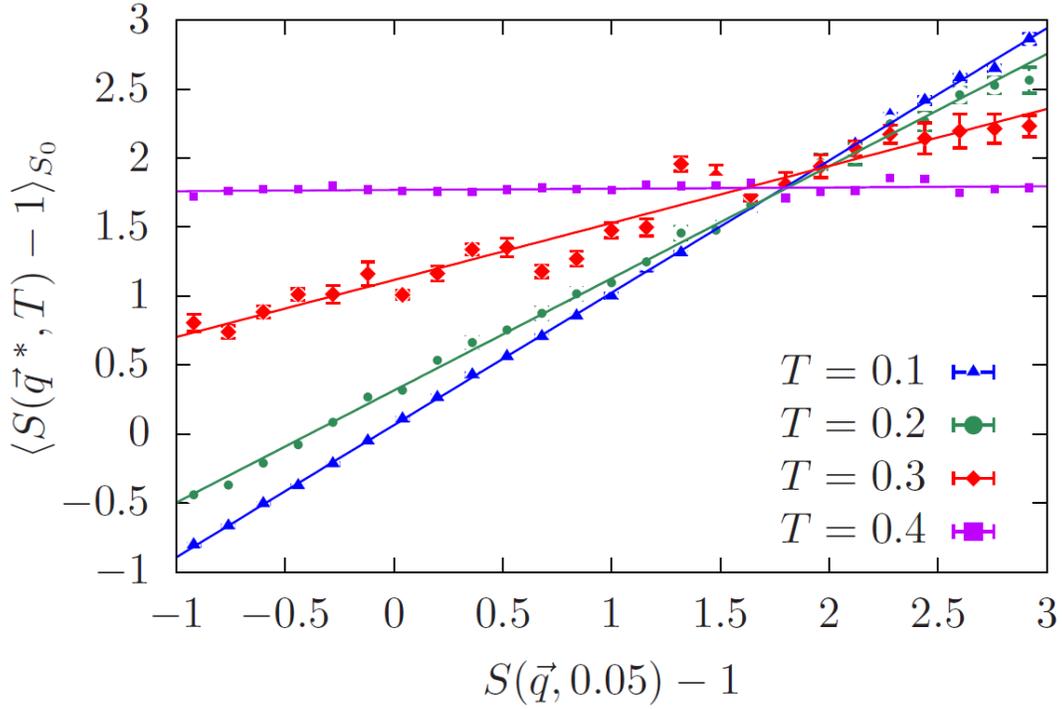

**Figure S6**. Plot of $\langle S(\vec{q},T)-1\rangle_{S_o}$ vs $S(\vec{q},0.05)-1$ at a constant pressure of 1.0 for 4 temperatures as indicated.

|  | $D_w(q_1,T)$ $T_{ref}=0$ | $D_w(q_1,T)$ $T_{ref}=0.05$ | $D_w(q_1,T)$ $P=0$ | $H(q_1,T)$ $T_{ref}=0$ | $H(q_1,T)$ $T_{ref}=0.05$ | $H(q_1,T)$ $P=0$ |
|---|---|---|---|---|---|---|
| T = 0.1 | 0.93 ±0.01 | 0.98±0.01 | 0.96±0.01 | 0.11±0.01 | 0.044±0.003 | 0.07±0.01 |
| T = 0.2 | 0.79±0.01 | 0.85±0.01 | 0.81±0.01 | 0.35±0.01 | 0.25±0.02 | 0.32±0.01 |
| T = 0.3 | 0.49±0.02 | 0.62±0.01 | 0.41±0.01 | 0.97±0.03 | 0.70±0.02 | 1.12±0.03 |
| T = 0.4 | 0.11±0.01 | 0.20±0.02 | 0.01±0.01 | 1.67±0.02 | 1.47±0.03 | 1.77±0.01 |

**Table S1** Comparison of values of $D_w(q_1,T)$ and $H(q_1,T)$ obtained using $T_{ref}=0$ and $T_{ref}=0.05$ (at fixed density) and for fixed pressure = 0 (with $T_{ref}=0.05$) for a range of temperatures.

To address the question of the impact of thermal expansion on the extraction of $D_w(q_1,T)$ and $H(q_1,T)$ we have carried out a set of simulations at a fixed pressure of 1.0 using a Nose-Hoover barostat with damping parameter of 20.0. The thermal expansion is accounted for by the selection of an appropriate reference q vector, $\vec{q}^* = \vec{q}_{T=0.05} \sqrt[3]{\dfrac{\rho(T)}{\rho(0.05)}}$ where ρ(T) is the density at temperature T. As shown in Fig. S6, we find a linear relation between $\langle S(\vec{q}^*,T)-1 \rangle_{S_o}$ and $S(\vec{q}^*,0.05)-1$. The extracted values for $D_w(q_1,T)$ and $H(q_1,T)$ are presented in Table S1.